\documentclass[conference]{IEEEtran}
\usepackage{cite}                       %��������
\usepackage{url}                        %������ַ
\usepackage{booktabs}                   %������õĺ��
\usepackage{diagbox}                    %������õĺ��
\usepackage{multirow}                   %����б����õĺ��
\usepackage{amsmath,amssymb,amsfonts,bm}
\usepackage{algorithmic}
\usepackage{graphicx}                   %��ͼƬ�ĺ��
\usepackage[subfigure]{graphfig}        %�����ͼƬ�ĺ��
\usepackage{textcomp}
\usepackage{xcolor}
\usepackage{verbatim}                   %����ע��
\usepackage{flushend}  
\usepackage{verbatim}
\usepackage{amsmath, amssymb}
\usepackage{makecell}
\usepackage{multirow}

\newcommand{\FF}{\mathbf{F}}
\newcommand{\RR}{\mathbf{R}}
\newcommand{\ZZ}{\mathbf{Z}}

\def\BibTeX{{\rm B\kern-.05em{\sc i\kern-.025em b}\kern-.08em
    T\kern-.1667em\lower.7ex\hbox{E}\kern-.125emX}}

\makeatletter
\def\Hline{%
	\noalign{\ifnum0=`}\fi\hrule \@height 1pt \futurelet
	\reserved@a\@xhline}
\makeatother

\begin{document}
	
	\title{SAG-GAN: Semi-Supervised Attention-Guided GANs for Data Augmentation on Medical Images}

	\author{\IEEEauthorblockN{Chang Qi$^{1,2}$, Junyang Chen$^{3}$, Guizhi Xu$^{1,2}$,Zhenghua Xu$^{1,2,\dag}$, Thomas Lukasiewicz$^{4}$ and Yang Liu$^{5}$}
		\IEEEauthorblockA{
			$^1$State Key Laboratory of Reliability and Intelligence of Electrical Equipment, Hebei University of Technology, China.\\
			$^2$Tianjin Key Laboratory of Bioelectromagnetic Technology and Intelligent Health, Hebei University of Technology, China. \\
			$^3$Department of Computer Science, University of Macau, China. \\
			$^4$Department of Computer Science, University of Oxford, United Kingdom. \\
			$^5$College of Computer Science and Technology, Harbin Institute of Technology, China. \\
			$^{\dag}$Corresponding author, email: zhenghua.xu@hebut.edu.cn}
	}

	\maketitle
	
	\begin{abstract}
        Recently deep learning methods, in particular, convolutional neural networks (CNNs), have led a massive breakthrough in the range of computer vision. Also, the large-scale annotated dataset is the essential key to a successful training procedure. However, it is a huge challenge getting such datasets in the medical domain. Towards this, we present a data augmentation method for generating synthetic medical images using cycle-consistency Generative Adversarial Networks (GANs). We add semi-supervised attention modules to generate images with convincing details.
        We treat tumor images and normal images as two domains. The proposed GANs-based model can generate a tumor image from a normal image, and in turn, it can also generate a normal image from a tumor image. Furthermore, we show that generated medical images can be used for improving the performance of ResNet18 for medical image classification. Our model is applied to three limited datasets of tumor MRI images. We first generate MRI images on limited datasets, then we trained three popular classification models to get the best model for tumor classification. Finally, we train the classification model using real images with classic data augmentation methods and classification models using synthetic images. The classification results between those trained models showed that the proposed SAG-GAN data augmentation method can boost Accuracy and AUC compare with classic data augmentation methods. We believe the proposed data augmentation method can apply to other medical image domains, and improve the accuracy of computer-assisted diagnosis.
	\end{abstract}
	
	
	\section{Introduction}
	
Over recent years, there is great progress in the field of Generative Adversarial Networks (GANs)\cite{goodfellow2014generative} and its extensions. Their wide and successful applications in image generation task\cite{mirza2014conditional,arjovsky2017wasserstein,radford2015unsupervised} have attracted growing interests across many communities, including medical imaging. GANs proposed a minimax game between two Neural Networks - generator generates samples, discriminator identifies the source of samples. In this game, the adversarial loss brought by discriminator provides a clever way to capture high dimensional and complex distributions, which imposed higher-order consistency that is proven to be useful in many cases, such as domain adaptation, data augmentation, and image-to-image translation.
	
    It is widely known that sufficient data is critical to success when training deep learning models for computer vision. Data with high class imbalance or poor diversity leads to bad model performance. This often proves to be problematic in the field of medical imaging, where abnormal findings are, by definition uncommon. While traditional data augmentation schemes (e.g., crop, rotation, flip, and translation) can mitigate some of these issues, those augmented images have a similar distribution to the original images, leading to limited performance improvement. Also, the diversity that those modifications of the images can bring is relatively small. Motivated by the GANs, researchers try to add synthetic samples to the training process. GANs-based data augmentation can improve performance by filling the distribution that uncovered by origin images. Since it can generate new but realistic images, it can achieve outstanding performance in medical image analysis\cite{lim2018doping,bowles2018gan,madani2018chest}.
	
	%One of the main challenges in the area of computer vision is how to cope with the small datasets' diversity and the limited amount of annotated samples, especially when applying supervised algorithms that require larger training samples. Many researchers attempt to overcome this challenge by using data augmentation schemes such as rotation, flip, scale and translation\cite{krizhevsky2012imagenet}. But the diversity that those modifications of the images can bring is relatively small. Motivated by the GANs, researchers try to use synthetic samples to improve the training process\cite{lim2018doping,bowles2018gan,madani2018chest}. 
	
	Despite these efforts, GANs based data augmentation in some topics of medical imaging such as converting a normal image without tumor to a tumor image remains a challenging problem. Essentially, this is an image generation problem, but unlike the style translation task such as synthesizing PET image from CT scans\cite{ben2017virtual} or MRI\cite{pan2018synthesizing}, the attribute manipulation is more challenging due to the requirement of only modifying some image features while keeping others unchanged. A straightforword option is to modify attribute manually in high dimension\cite{shin2018medical}, convert the attribute manipulation task to style transform task. The other option is to generate tumor images from noise\cite{han2019learning} rather than normal images. However those solutions define the position and size of tumors manually, which may break image prior, cause unacceptable additional false positives in the following image processing tasks.        
	
To overcome the issues mentioned above, in this paper, we proposed a novel GAN-based data augmentation model guided by semi-supervised attention mechanisms. Inspired by CycleGAN\cite{zhu2017unpaired}, The proposed data augmentation model comprises of two generators and two discriminators. The cycle-consistence constrains the model to change the image features that need to be modified. Besides, It not only generates images by enabling manipulate high-level features but also pushes the generator to locate the areas to translate in each image with additional attention module. The attention modules are trained by both adversarial loss and pixel-wise loss. That is why we call it semi-supervised attention mechanisms. Moreover, we add spectral normalization\cite{miyato2018spectral} to stabilize the training of the discriminator. We evaluate the generated images' realism by tumor classification results with/without the proposed data augmentation method.
	
	The contributions of this paper are summarized as follows:
	
	\begin{itemize}
		\item \textbf{GAN-based data augmentation method for medical image:} We proposed a novel semi-supervised attention-guided CycleGAN to generate tumor in normal images and recover normal images from tumor images. It's the first model that generate tumor in normal images naturally. So that the image prior will not be destroyed.%This semantic manipulation method can be used in other applications in computer vision, besides this attention supervised generator and discriminator can be used in other GANs models easily.
		\item \textbf{Semi-supervised attention mechanism:} To the best of our knowledge, we are the first that integrates the semi-supervised attention mechanism into Generative Adversarial Networks. We proposed an attention module trained with adversarial loss and pixel-wise loss. The additional pixel-wise loss push the attention module locates the location of tumors as accurately as possible.
	\end{itemize}
	
	\section{Related Work}
    \textbf{Generative Adversarial Networks(GANs)}\cite{goodfellow2014generative} have two models to train: a model $G$ to learn the target data distribution $p_{\text{data}}(\bm{x})$, a model $D$ to assess the source of $D$'s input, is it from $p_{data}(x)$ or from model $G(\bm{z})$. The aim of the training model $G$ is to maximize the chance of model $D$ making mistakes, while the aim of training model $D$ is to maximize the probability of assigning the correct label to both training examples and samples from $G$. GANs are powerful generative models, which have achieved impressive results in many computer vision tasks. The adversarial loss is the key to GAN's success. It forces the model to generate images that indistinguishable from real images. The optimism of the generator is not the pixel-wise loss but another network-discriminator.
	
	%it defines a prior on input noise variables $p_z(\bm{z})$, then represent a mapping to data space as $G(z;\theta_g)$, $\theta_g$ are parameters in function $G(\bm{z})$;, function $D(\bm{x})$ is defined to represents the probability that $\bm{x}$ ($D$'s input) came from the data rather than $G(\bm{z})$.  
	
    \textbf{Attention-guided GANs} address the issues that the instance-level correspondences are indistinguishable from the distribution of the target set. The attention module forces the network to pay more attention to the area under the attention map. 
Mejjati \textit{et al.} \cite{mejjati2018unsupervised} propose an attention mechanism that is jointly trained with the generators and discriminators.
Chen \textit{et al.} propose AttentionGAN \cite{chen2018attention}, an extra attention network is used to generate attention maps.
Kastaniotis \textit{et al.} \cite{kastaniotis2018attention} proposed ATAGAN, a teacher network is used to produce attention maps.
Zhang \textit{et al.} \cite{zhang2018self} proposed a Self-Attention Generative Adversarial Networks (SAGAN). The Non-local module\cite{wang2018non} is used to produce the attention map. 
Liang \textit{et al.} \cite{liang2017generative} propose a Contrasting GAN that takes the segmentation mask as the attention map. 
Sun \textit{et al.}~\cite{sun2018mask} proposed an attention GANs using FCN to generate a facial mask for face attribute manipulation.
	
	% generate abnormal MRI images with brain tumors from normal MRI images.
To the best of our knowledge, there are two research groups tried to generate tumor images. Shin et al.\cite{shin2018medical} tried to duplicate tumors in BRATS dataset to normal MRI images in ADNI dataset. They proposed a two-stages GANs model for the transform task. The first stage is an image-to-brain segmentation GAN-based model. The generator generates the brain masks with white matter, grey matter, and CSF from the input MRI images. The discriminator is trained to distinguish the real brain masks that annotated by doctors versus the fake brain masks generated by the generator. The second stage is a brain-to-image synthesis GAN-based model. The brain masks generated in stage one was merged with tumor masks. The merged brain masks as the brain-to-image model's input generate the abnormal MRI images with brain tumors. However, there are some limits in this two-stages model. Firstly, there is no brain mask annotation in dataset A. Thus the tumor mask that merged to brain mask is inferred by the model trained by dataset B. But dataset B does not contain tumor information, so the quality of the generated tumor mask is doubtful. Secondly, the position of the tumor merged to brain mask is decided by researchers, which means it is manually and random. However, the tumor location is related to other feature of the tumor, such as size, shape, degree of malignancy, the attempt that locate a wrong tumor location with specific other features may damage the MRI image prior, causing higher false positives and less robustness of the model.
	
 Han et al.\cite{han2019learning} proposed a CPG-GAN model for generating tumor in the normal MRI images. The 'condition' in their model is a $[0,1]$ mask, where $0$ stands for the nontumor area, $1$ stands for the area that needs to generate tumor feature. However, there are some problems in there work: Firstly, like shin et al.'s work\cite{shin2018medical}, the position of the tumor is decided by researchers, which will damages the image prior. The experiments in work prove it: the FPs per slice of detection task increase $3.52$ with only $0.1$ increase of sensitivity. Secondly, the adversarial loss is not enough to generate a realistic tumor image from normal MRI image. 

	% \section{Our approach}
	% 	\begin{figure}[t]
	% 	\includegraphics[width=\linewidth]{./PICS/III/III_1.pdf}
	% 	\centering
	% 	\label{fig_architecture}
	% 	\vspace{-5mm}
	% 	\caption{\small The framework of the proposed data augmentation model. 'T' represents for Tumor samples, 'N' represents for Normal samples. G1/G2: The Attention-guided generator, D1/D2: The Attention-guided discriminator.}
	% 	\vspace{-5mm}
	% \end{figure}

	\begin{figure*}[h]
		\includegraphics[width=\linewidth]{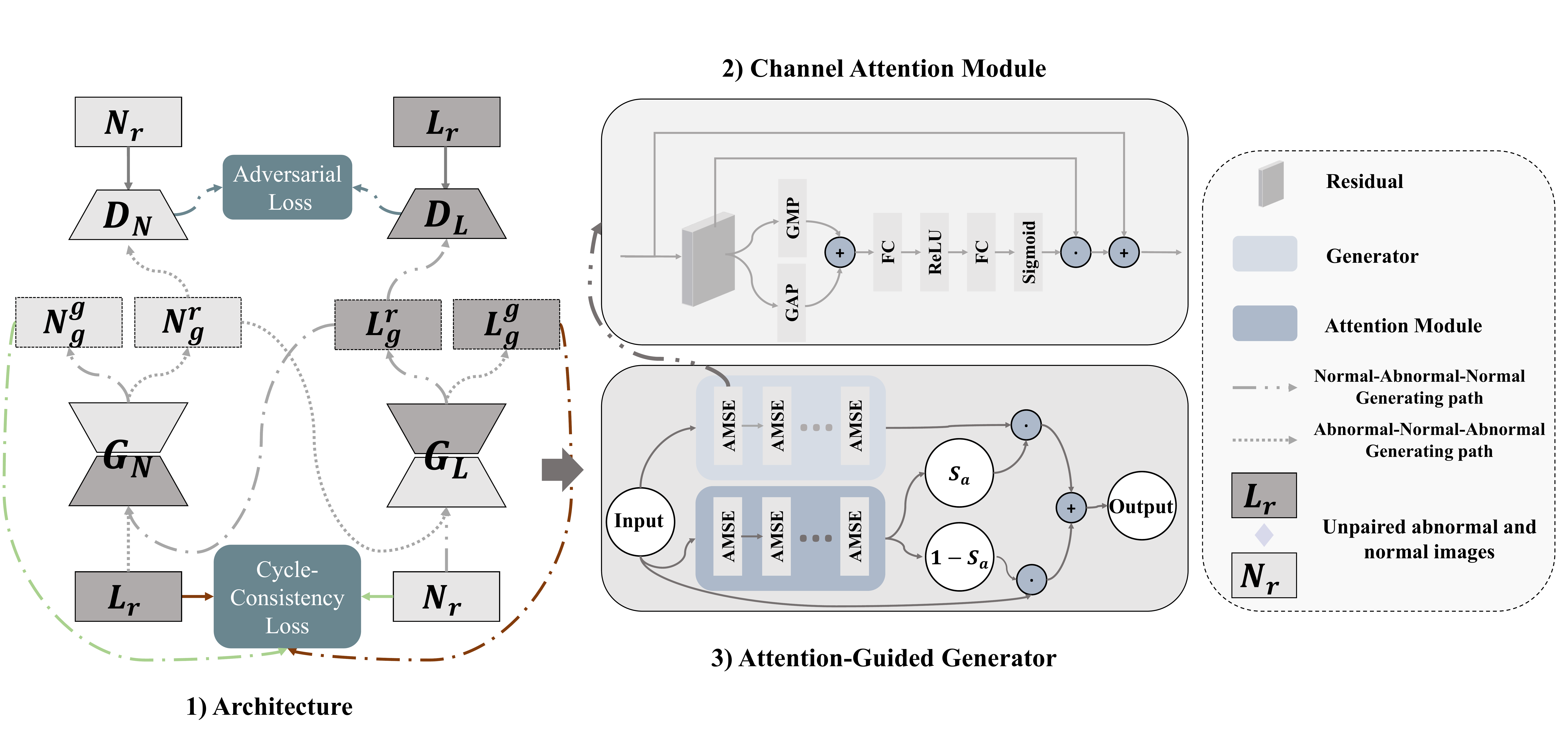}
		\centering
		\label{wm}
		\vspace{-5mm}
		\caption{\small Illustration of our proposed Semi-supervised Attention-Guided CycleGAN (SAGGAN). 'L' represents for abnormal images with target object, 'N' represents for normal images. $G_L$ / $G_N$: The Attention-guided generators, $D_L$ / $D_N$: The Attention-guided discriminators.}
		\vspace{-5mm}
	\end{figure*}
	
    The goal of our work is to generate abnormal MRI images from normal MRI images to fulfill the data distribution. It means that the model needs to learn the mapping function from normal domain ($N$) to tumor domain ($T$) with unpaired training samples $\{N_i\}^{n}_{i = 1}\in N$ and $\{T_j\}^{m}_{j = 1} \in T$. As illustrated in Fig. 1, our data augmentation method includes two mapping function G1: Normal$\rightarrow$Tumor and G2: Tumor$\rightarrow$Normal. Meanwhile, there are two adversarial discriminators D1: distinguish between real tumor images $\{T_r\}$, generated tumor images from real normal images $\{T_g^r\}$, and generated tumor images from generated normal images $\{T_g^g\}$, in the same way, D2: distinguish between real normal images $\{N_r\}$ , generated normal images from real tumor images $\{N_g^r\}$, and generated normal images from generated tumor images $\{N_g^g\}$. The attention networks in both generator G1/G2 and discriminator D1/D2 predict the region of interest in the input image.

	%To overcome the above problem, CycleGAN approch contrains two mapping functions: $F_{S \rightarrow T}$ and $F_{T \rightarrow S}$, the mapping function $F_{T \rightarrow S}$ is designed to constrain the optimizime path by add cycle consistency loss: $\lVert X_S - F_{T \rightarrow S}(F_{S \rightarrow T}(X_S))\rVert_1$, the training stategy of network $F_{S \rightarrow T}$ requires both the discriminator $D_T$ to detect the generate sample $F_{S \rightarrow T}(X_S)$ from the real sample $X_T$ and reduce the cycle consistency loss. For cycle consistance, the inverse generator $F_{T \rightarrow S}$ and discriminator $D_S$ are trained simultaneously.
	
\section{Proposed Model}	
\subsection{Channel Attention Module}
\label{section:ca}
%\begin{figure}
%	\includegraphics[width=2.5in]{./figures/AMSE.pdf}
%	\centering
%	\caption{\small The framework of the AMSE-Residual Module.}
%	\label{amse}
%\end{figure}
The residual block built the whole model, to enhance the model ability in capture hierarchical patterns and then improve the quaility of image representations, we adopt improved Squeeze-and-Excitation(SE) block\cite{hu2018squeeze} to the original residual module following the standard intergration design. The goal is achieved by explicitly modeling the interdependencies between the channels in feature space, which allows the model to emphasise channels with informative features and suppress less useful channels. 

Specifically, as shown in Fig.~\ref{wm}, the proposed AMSE (Average-Maximum Squeeze-and-Excitation) -Residual module consistent with two blocks: the transformation function $\FF _t$, and the AMSE block. $\FF _t$ takes the image features $X \in \RR^{C' \times H'\times W'}$ captured in the last AMSE-Residual block as input, and outputs transformed image features $U \in \RR^{C\times H \times W}$. At the squeeze stage, a statistic $\ZZ \in \RR^C$ is generated by ignore U's spatial dimensions $H \times S$. The $p_{th}$ element of $\ZZ$ is calculated by:
\begin{equation}\label{squeeze}
Z(p) = \FF_{sq}(U) = \frac{1}{H\times W}\sum_{i=0}^{H-1} \sum_{j=0}^{W-1} U(p,i,j) + \max U(p,i,j)
\end{equation}

To learn the nonlinear interaction between channels while maintain the non-mutually-exclusive relationship, at the excitation stage, a gating mechanism consist with two fully-connected (FC) layer and one relu layer is being used. Given the output of the squeeze stage $\ZZ$, the output activate $\ZZ' \in \RR^C$ of the excitation stage is:
\begin{equation}\label{excitation}
\ZZ' = \FF_{ex}(\ZZ)
\end{equation}

The final outputs $X' \in \RR^{C \times H\times W} $of this block is the rescaled $U$ plus the origin $X$:
\begin{equation}\label{rescale}
X = \FF_{rescale}(\ZZ',U) + X = \ZZ' U + X
\end{equation}

	\subsection{Attention-Guided Generator}
	
    Unlike style translation tasks in domain translation tasks, translation between normal images to tumor images require to solve two tasks: 1) location the area to translate, and 2) taking the proper translation in the located area. So we proposed two attention networks $A_N$ and $A_T$ to achieve this. Where $A_N$ aims to select the area to generate tumor that maximizes the probability that the discriminator makes a mistake and minimizes the probability that the generator makes a mistake; $A_T$ aims to locate the place that has tumor and generates the possibility map, which will guide the generator recover the normal images from tumor images.
	
     In the forward processing, the generated image contains two parts, the foreground from the generator and the background from the input image. %Firstly, the input image fed into the generator will element-wise with the corresponding attention mask $S_a$ that contains per-pixel [0,1] estimates. Secondly, the inverse of the attention map will apply to the input image as the background. Thirdly, the above two parts will be element-add as the generated image. 
Take the translation from normal samples to tumor samples as an expmple. Firstly, the normal brain MRI $\{N_r\}\in N$ is fed into the generator $G_{N \rightarrow T}$, which maps $\{N_r\}$ to the target domain $T$, resulting the generated tumor image $G_{T'} = G_{N \rightarrow T}(\{N_r\})$. Then, the same input $\{N_r\}$ is fed into the attention module $A_N$, resulting in the attention map $M_{N'} = A_N(\{N_r\})$.
To create the `foreground' object $\{{T'_f}\}\in T$, we apply $M_{N'}$ to $G_{T'}$ via an element-wise product: $\{{N'_f}\} \!=\! M_{N'}\odot G_{T'}$ .
Secondly,  the inverse of attention map $M_{N'}' = 1-M_{N'}$ will be applied to the input image via an element-wise product as the background.
Thus, the mapped image$\{{T_g}\}$ is obtained by:
	\begin{equation}
	T_g = \underbrace{M_{N'}\odot G_{T'}}_\text{Foreground}  +  \underbrace{ M_{N'}' \odot N_r}_\text{Background} \text{.} % 
	\label{e:one}
	\end{equation}
	
	We only described the map $F_{N \rightarrow T}$; the inverse map $F_{T \rightarrow N}$ is defined similarly. Fig. 2  visualizes those processes.

	%But there are a little difference in the training processing of two mapping function. As we mentioned before, locating the area to manipulation plays a key role in the network, and the attention network $A_S$, $A_T$ is proposed to solve the problem. But vanilla adversatial loss and cycle-consistency loss is not enough for the optimization of attention module. So we denote the brain MRI image with tumor as $t \in T$ with semantic $M_T$. As illustarted in Figure ?, the $A_T$ module is optimizated by vanilla adversarial loss, cycle-consistency loss and pixel-loss.

	\subsection{Attention-Guided Discriminator}
	
    Eq.~\ref{e:one} constrains the generators to modify only on the attention regions: 
However, the discriminators currently consider the whole image. Vanilla discriminator $D_T$ takes the whole generated image $\{T_g\}$ and the whole real image $\{T_r\} \in T$ as input and tries to distinguish them.  
The attention mechanism is added into the discriminator so that discriminators only consider the regions inside the attention map.
We propose two attention-guided discriminators.
The attention-guided discriminator takes the attention mask, the generated images, and the real images as inputs.
For attention-guided discriminator $D_{T}^A$, which tries to distinguish the fake image with attention map $M_{N'}\odot T_g$ and the real image with attention map $M_{N'}\odot T_r$.
Similar to $D_{T}^A$, $D_{N}^A$ tries to distinguish the fake image with attention map $M_{T'}\odot N_g$ and the real image with attention map $M_{T'}\odot N_r$.
Discriminators can focus on the most discriminative content by this attention-guided method.
	
	\subsection{Spectral Normalization}
    It is wildly known that the reason why GANs is challenging to train is that the objective function of the native GANs is equivalent to the J-S divergence between the distribution $p_g$ of the optimized generated data and the distribution $p_r$ of the real data. However, J-S metric fails to provide a meaningful value when two distributions are disjoint. It makes no guarantee convergence to a unique solution such that $p_r = p_g$. Then, WGAN\cite{arjovsky2017wasserstein} was proposed to replace the J-S divergence in the native GAN with a good Wasserstein distance. The KR duality principle is used to transform the Wasserstein distance problem into a solution to the optimal Lipschitz continuous function. The spectral normalization proposed by Miyato et al.\cite{miyato2018spectral} use a more elegant way to make the discriminator meet Lipschitz continuity.
	
	\subsection{Semi-supervised Attention Mechanism}
Segmentation annotations are available in our case. For example, the attention map of tumor $\rightarrow$ normal translation is exactly the whole tumor region in the segmentation annotation of the tumor. Therefore, we supervise the training process of attention network $A_T$ by segmentation label. Given a training set $\{(T_1,M_1),\cdots,(T_N,M_N)\}$ of $N$ examples, where $M_i$ stands for the tumor label of segmentation. To reduce changes and constrain generators, we adopt pixel-wise loss between the tumor label $M_i$ and the generated attention map $M_{T_i'}$.
We express this loss as:
	\begin{equation}
	\begin{aligned}
	\mathcal{L}_{M}(M_i,M_{T_i'}) = \Arrowvert M_i - M_{T_i'}\Arrowvert_1.
	\end{aligned}
	\label{equ:pixelloss}
	\end{equation}
	
	%\begin{comment}

	We adopt $L1$ distance as loss measurement in pixel loss.
	Also, we denote the tumor segmentation label as the ground-truth annotation for the attention map $M_{T_i'}$.
	This added loss makes our model more robust by encouraging the attention maps to be sharp (converging towards a binary map), while the attention mask of normal areas will always be zero.

	\subsection{Optimization Objective}	
	Fig. 3 shows the training loss of the proposed data augmentation model.
	
	\begin{figure}[h]
  		\includegraphics[width=\linewidth]{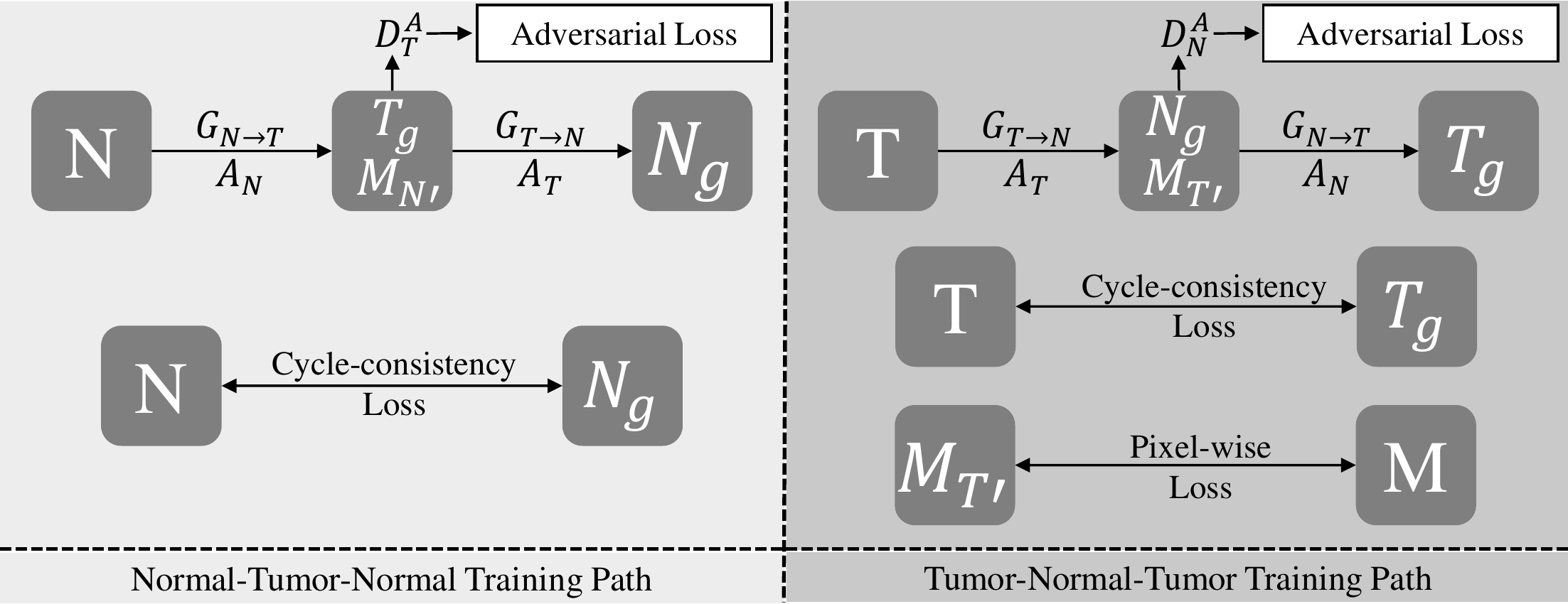}
		\centering
		\label{fig_loss}
		\vspace{-5mm}
		\caption{\small The training loss of the proposed GAN-based network. }
		\vspace{-3mm}
	\end{figure}
	\noindent \textbf{Attention-guided Adversarial Loss} The attention-guided adversarial loss is proposed to training the attention-guided discriminators. It can be formulated as follows:
	\begin{equation}
	\begin{aligned}
	\label{eq:minimax}
	\mathcal{L}_{AGAN}^{N}(&G_{N\rightarrow T}, D_T^A) =  \mathbb{E}_{t\sim{p_{\rm data}}(t)}\left[ \log D_T^A(M_{N'}\odot t)\right] + \\
	& \mathbb{E}_{n\sim{p_{\rm data}}(n)}[\log (1 - D_T^A(M_{N'}\odot G_{N\rightarrow T}(n)))].
	\end{aligned}
	\end{equation}
	where $ G_{N\rightarrow T}$ aims to translation the normal image to tumor image and maximize the probability that the discriminator makes a mistake, while $D_T^A$ trained to distinguish between the generated image with its attention mask
    $(M_{N'}\odot t)$. 
    Which means  $ G_{N\rightarrow T}$ 
    tries to minimize the attention-guided adversarial loss 
    $\mathcal{L}_{GAN}(G_{N\rightarrow T}, D_T^A)$, 
    while $D_T^A$ tries to maximize it. There are also another loss for the discriminator $D_N^A$ and the generator $ G_{T\rightarrow N}$:
    \begin{equation}
    \begin{aligned}
    \label{eq:minimax2}
    \mathcal{L}_{AGAN}^{T}(&G_{T\rightarrow N}, D_N^A) =  \mathbb{E}_{n\sim{p_{\rm data}}(n)}\left[ \log D_N^A(M_{T'}\odot n)\right] + \\
    & \mathbb{E}_{t\sim{p_{\rm data}}(t)}[\log (1 - D_N^A(M_{T'}\odot G_{T\rightarrow N}(t)))].
    \end{aligned}
    \end{equation}
	
	\noindent \textbf{Cycle-Consistency Loss.} 
	The cycle-consistency loss can be used to enforce forward and backward consistency.
	For example, if a tumor image is transformed into a normal image, the transformed from the generated normal image to the tumor image should be brought back to a cycle.
	Thus, the loss function of cycle-consistency is defined as:
	\begin{equation}
	\begin{aligned}
	&  \mathcal{L}_{cycle}(G_{N\rightarrow T}, G_{T\rightarrow N}) = \\
	& \mathbb{E}_{n\sim{p_{\rm data}}(n)}[\Arrowvert G_{T\rightarrow N}(G_{N\rightarrow T}(n))-n\Arrowvert_1]  + \\
	& \mathbb{E}_{t\sim{p_{\rm data}}(t)}[\Arrowvert G_{N\rightarrow T}(G_{T\rightarrow N}(t))-t\Arrowvert_1].
	\end{aligned}
	\label{equ:cycleganloss}
	\end{equation}

	\noindent \textbf{Loss Function.}
    We obtain the final energy to optimize by combining the adversarial loss, cycle-consistency loss, and semi-supervised pixel losses for both source and target domains:
	\begin{equation}
	\begin{aligned}
	\label{e:objective}
	& \mathcal{L}(G_{N \to T}, G_{T \to N}, A_N, A_T, D_N^A, D_T^A) =  \\
	& \lambda_{gan} (\mathcal{L}_{AGAN}^N + \mathcal{L}_{AGAN}^T) + \\
	& \lambda_{cyc} \times \mathcal{L}_{cycle}(G_{N\rightarrow T}, G_{T\rightarrow N}) + \mathcal{L}_{M}(M_i,M_{T_i'})
	\end{aligned}
	\end{equation}
	
	%\end{comment}

	\section{Experiments and Results}

    We present sets of experiments and results in this section. To evaluate the performance of proposed CycleGAN-based data augmentation method, we employed a convolutional neural network with the deep residual block (ResNet18)\cite{he2016deep} 
    to compare the classification results using generated tumor images to the classification results of real images. We implemented five models to generate tumor images, as described in Section \textrm{IV.A.b}.
	
	For the implementation of the tumor classification model ResNet18 and GAN-based data augmentation architecture, we used the Pytorch framework. All training processes were performed in an NVIDIA GeForce GTX 1080 Ti GPU.
	
	\subsection{Dataset Evaluation and Implementation Datails}
	\subsubsection{Classification}
    For brain tumor classification, we chose ResNet18\cite{he2016deep} because, among three popular neural network models\cite{simonyan2014very,krizhevsky2012imagenet,he2016deep}, our initial experiments showed that ResNet18 % took short training time, best classification performance and fewer parameters is the best model in the tumor classification task. %Besides, it reduces the risk of overfitting. 
%We compared the accuracy of three neural network architectures on three datasets, Resnet18 
took the shortest training time, the best classification performance and has the lowest number of parameters, making it potentially more portable and less prone to overfitting. In this study, we split the datasets into 70\% training, 20\% validation, and 10\% test images.

We calculated TPR and TNR to meature the performance of our data augmentation method in tumor classification task. In the following equations, we present these measures:

	\begin{equation}
	TPR = \frac{TP}{TP + FN}
	\end{equation}
	
	\begin{equation}
	TNR = \frac{TN}{TN + FP}
	\end{equation}
	
    where $T$ stands for correct classification, in the opposite, $F$ stands for the wrong classification. $P$ stands for the classification result is tumor category, and $N$ stands for the classification result is the normal category. So, $TP$ stands for the tumor image is classification to the tumor category, $TN$ stands for the normal image is classification to the normal category, $FP$ stands for the normal image is classification to the tumor category, and $FN$ stands for the tumor image classification to the normal category.

	\subsubsection{Baselines} We compare our model with leading image to image translation model: CycleGAN\cite{zhu2017unpaired}, and the extension of CycleGAN: Attention-guided CycleGAN\cite{mejjati2018unsupervised}. For a fair comparison, we then add the spectral normalization method to those models.
	
	\begin{comment}
			\textbf{CycleGAN\cite{zhu2017unpaired}}: This method first introduce the Cycle-Consistency loss into the GANs framework, it learns the translation from domain X to domain Y while learns the translation from domain Y to domain X with another generator. 
		
		\textbf{Spectral norm.\cite{miyato2018spectral} + CycleGAN}: We also add spectral normlization to CycleGAN to 
		
		\textbf{Attention + CycleGAN\cite{mejjati2018unsupervised}}:Like our method, Mejjati et al. uses an attention module to train the translation from source domain to target domain.
		
		\textbf{Spectral norm.\cite{miyato2018spectral} + Attention CycleGAN}:We also add spectral normlization to Mejjati's attention CycleGAN.
	\end{comment}
	
	\subsubsection{Datasets} We use the BraTS dataset provided by Menze et al.\cite{menze2014multimodal} to evaluate our data augmentation method. These datasets contain segmentation mask for each case, and for each case, there are four modal MRI images: T1, T1c, T2, and Flair. 
	
	There are 322 cases in the Brats 2019 dataset, where 20\% of all cases is signed as the testing data, and 10\% of all the cases if signed as the validation data. To test the performance of the proposed data augmentation scheme with limited data, we randomly selected one-eighth of the training data as the training set $\text{BraTS}_S$.
	
%	\subsubsection{Data processing and Training details} 
	
	\subsection{Evaluation of the Data Augmentation}
	\subsubsection{Generated Images by proposed SAG-GAN}
	\begin{figure}[h]
	%\vspace{-3mm}
	\includegraphics[width=\linewidth]{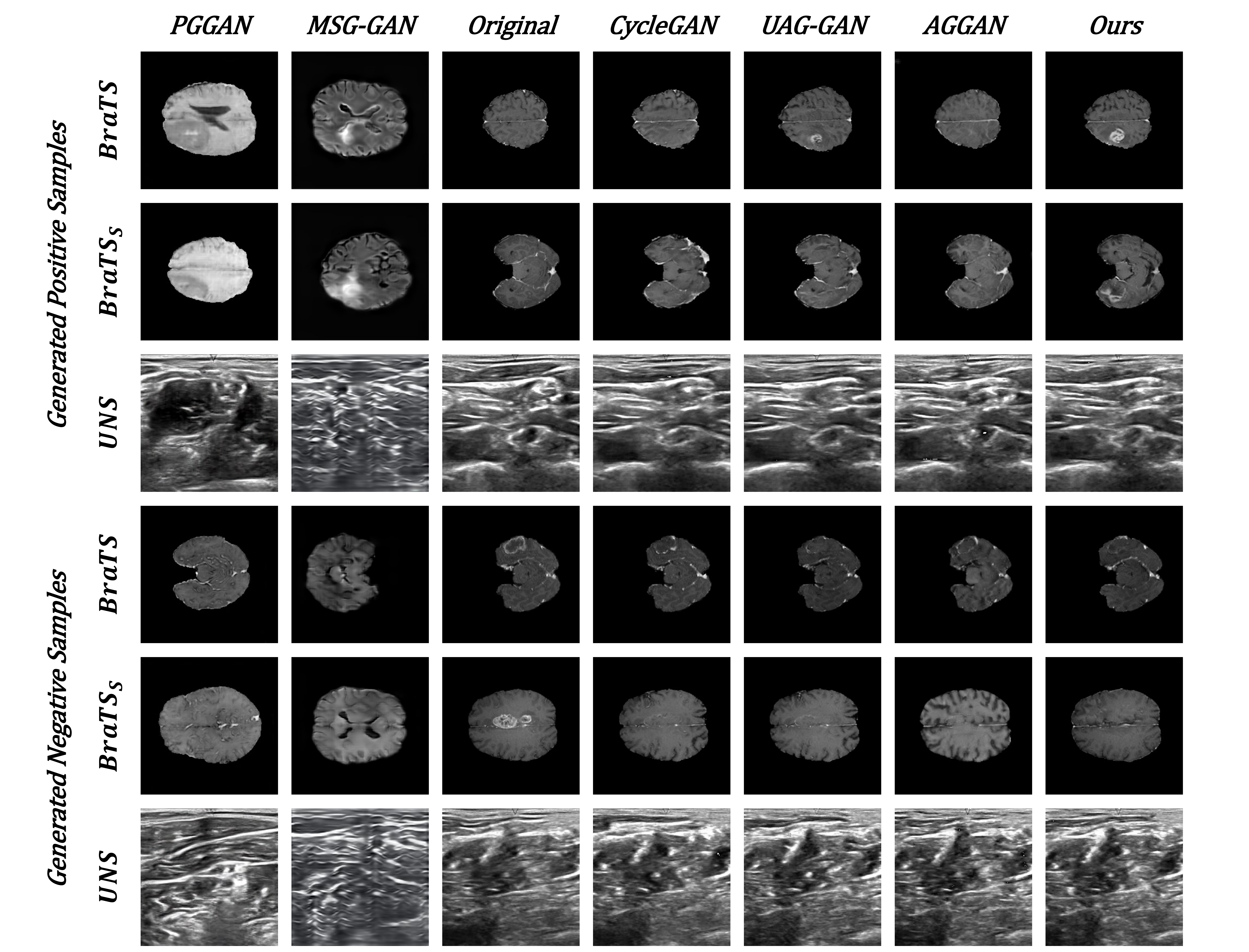}
	\centering
	\label{VI_1}
	\vspace{-5mm}
	\caption{\small Results of synthtic images.}
	\vspace{-3mm}
\end{figure}
    Fig. 4 illustrates examples of synthetic MR images by our data augmentation method. Observing the generated images and learned attention maps by our model, we can see that our model successfully captures the T1c-specific texture and tumor appearance in the right position. 
	\subsubsection{Quantitative results}
	\begin{table*}
		\centering
		\caption{Comparison with State-Of-The-Arts}
		\small
		\begin{tabular}{c|ccc|ccc}
			\noalign{\smallskip}\toprule
			& \multicolumn{3}{c|}{Accuracy} & \multicolumn{3}{c}{AUC} \\
			\midrule
			Method & BraTS  & $\text{BraTS}_S$ & UNS & BraTS & $\text{BraTS}_S$ & UNS\\ 
			\midrule
			w/o DA & 0.9337& 0.8334 & 0.6837& 0.9612&0.9306	&0.7940\\
			OverSampling & 0.9375 & 0.8418 & 0.7196 & 0.9643 & 0.9328 &	0.8072\\
			UnderSampling & 0.9369 & 0.8373& 0.7285&0.9647 & 0.9314 & 0.8192 \\
			\midrule
			PGGAN\cite{karras2017progressive} & 0.9419 & 0.8500 & 0.7331 & 0.9670 & 0.9345 & 0.8199\\
			MSG-GAN\cite{karnewar2019msg} & 0.9410 & 0.8530 & 0.7354 & 0.9667 & 0.9350 & 0.8190 \\
			\midrule
			CycleGAN\cite{zhu2017unpaired} & 0.9430 & 0.8539& 0.7387 &0.9679&0.9354&0.8208	\\
			UAGGAN\cite{mejjati2018unsupervised} & 0.9467&	0.8704&	0.7634 &0.9689	&	0.9465&	0.8290 \\
			AGGAN\cite{tang2019attention} & 0.9460&	0.8587&	0.7457& 0.9682&	0.9353&	0.8226\\
			Ours & \textbf{0.9503}&	\textbf{0.8759}&\textbf{0.7650}&\textbf{0.9697}	&\textbf{0.9530}	&\textbf{0.8297}\\
			Relative imp.\% &+25.03\% &+25.51\% &+25.70\% &+21.91\%&+32.27\% &+17.33\% \\
			
			\toprule
		\end{tabular}
		\label{tab_resnetpre}
	\end{table*}
    Table.~\ref{tab_resnetpre} shows the performance of our data augmentation method in the image classification task. 
The results in Table.~\ref{tab_resnetpre} proved our hypothesis that adding generated samples can improve the classification performance. 

	\section{Conclusions}
    This work focus on generating abnormal images from normal images and recovering normal images from abnormal images with GAN. This GAN-based data augmentation method can enlarge small medical datasets, fulfill data distribution. While recent data augmentation method in the medical image can generate a new abnormal sample, they also have some limits. For example, previous works need masks to teach the generator the place to generate abnormal lesion. Also, it is hard for GANs-based model to generate a large-size medical image. Most generated abnormal samples are small ($32px \times 32px$). 
    The data augmentation method we proposed can generate abnormal images of the real medical image size (In this case, $240px \times 240px$). We expect to get significant improvements in the quality of generated abnormal images by incorporating an attention module into both generator and discriminator. However, then we found that the attention mapping is not robust since the shape of abnormal lesion is changing between images. So we add a semi-supervised mechanism to stabilize this training procedure. The result shows that this approach improves the robust of attention modules. Experimental results on three datasets demonstrate that our data augmentation method can generate striking results with convincing details than the state-of-the-art models.

    There are several limitations to this work. One possible extension could be the evaluation from the classification task to the segmentation task. Data insufficient happens more in the field of tumor/lesion segmentation. In the future, we plan to extend our work to other medical domains that can benefit from generated abnormal images to improve training performance.

\section*{Acknowledgments}
\noindent This work was supported in part by the National Natural Science Foundation of China under grant 61906063, in part by the Natural Science Foundation of Tianjin, China, under grant 19JCQNJC00400, and in part by the Yuanguang Scholar Fund of Hebei University of Technology, China.

	%�ο�����
	\bibliographystyle{IEEEtran}
	\bibliography{cite}

\end{document}